# Efficient short-wave infrared upconversion by self-sensitized holmium-doped nanoparticles

Rakesh Arul[1], Zhao Jiang[1], Xinjuan Li[2], Fiona M. Bell[1], Alasdair Tew[1], Caterina Ducati[2], Akshay Rao[1*] and Zhongzheng Yu[1*]

[1] Cavendish Laboratory, University of Cambridge, Cambridge, CB3 0HE, United Kingdoms
[2] Department of Materials Science and Metallurgy, University of Cambridge, Cambridge, CB3 0FS, United Kingdoms

*E-mail address: ar525@cam.ac.uk (A. Rao), zy338@cam.ac.uk (Z. Yu)

Photon upconversion, combining several low-energy photons to generate one high-energy photon is of wide interest for biomedical, catalytic and photonic applications. Lanthanide-doped nanoparticles (LnNP) are a unique type of upconversion nanoconverter, which can realize ultralarge anti-Stokes shift (>1000 nm) and high photostability, without photo-bleaching and photo-blinking. The excitation wavelength of LnNPs has been limited to the second near-infrared window (1000-1700 nm), mainly sensitized by erbium ions with absorption centered around 1.5 µm. Here, we demonstrate novel self-sensitized holmium (Ho)-doped nanoconverters to further expand the sensitization range to the short-wave infrared at 2 µm and achieve efficient upconversion to 640 nm. We show that this upconversion is a 4-photon conversion process with an underlying energy transfer upconversion mechanism. Via careful control of dopant concentration and shelling we achieve a relative upconversion-to-downconversion efficiency up to 15.2%, more than half the theoretical maximum. The placement of the Ho doped LnNPs into a plasmonic nanocavity device enables large gains in emission intensity (up to 32-fold), due to the dramatic shortening of the emission lifetime of Ho from 29 µs to <1 ns, indicating a high Purcell-enhancement factor of $3 \times 10^4$. These results open new possibilities at the frontier of short-wave infrared upconversion and the nanoplasmonic enhancement of LnNP emission, with potential applications in detection, theranostics, photonics and optoelectronics.

Upconversion is a nonlinear photon conversion process that has been applied in light-controlled release[1], photodynamic therapy[2], optogenetics[3,4], multiplexing[5,6], photocatalysis[7], anti-counterfeiting[8], 3D printing[9], detection[10,11], and super-resolution nanoscopy[12]. Among the different kinds of upconversion nanoconverters, lanthanide-doped nanoparticles (LnNPs) show many unique properties, such as high photo-stability, low cytotoxicity, ultralong anti-Stokes shift and tunable emission wavelength and lifetime[13]. Current research has explored the upconversion excitation range of LnNPs to the second near-infrared (NIR-II, 1000-1700 nm) range[5,14], mainly via the sensitization of $Er^{3+}$ ions at 1530 nm. Further expansion of the excitation range to the lower energy infrared regime remains an open problem.

Short-wave infrared (SWIR, 1 to 2.5 µm) light is critical to modern photonic technology. Behind all the applications of SWIR light are two key properties: low attenuation except within certain absorption bands, and lower Rayleigh scattering than visible light. Hence, SWIR provides strong contrast high-resolution imaging[15], with better penetration through fog and reduced scattering in deep tissue. SWIR light also experiences less thermal background noise compared to mid-infrared light, acting as a 'sweet spot' for chemical imaging. SWIR can also be used in other applications, including but not limited to telecommunications, space and remote sensing[16]. Despite the many advantages of SWIR light, highly sensitive and low-cost Si based semiconductor photodetectors cannot detect SWIR light as the photon energies are below the Si bandgap. This necessitates the use of expensive, high dark noise photodetectors based on semiconductors such as InGaAs, HgCdTe, or PbS. Improving the efficiency of such detectors often requires cooling, due to the presence of a thermal background. The direct upconversion of SWIR to the visible or NIR-I (700-1000 nm) would be a straightforward way to harvest SWIR and take advantage of established silicon photodetector technology. However, the direct upconversion of SWIR light has barely been reported. Previous attempts to upconvert light in the SWIR and MIR require multiple lanthanide energy transfers[17,18], or dual visible and MIR light excitation[19,20], reducing its simplicity, efficiency, and practicality.

Here, we design novel holmium ($Ho^{3+}$)-based nanoconverters to directly absorb SWIR at around 2 µm and efficiently perform upconversion to generate red emission at 640 nm. The design of this system is based on the transition of $^5I_8$ to $^5I_7$ electronic levels in $Ho^{3+}$ ions and direct self-sensitization to achieve upconversion. Figure 1a shows the core-shell structural design of these nanoconverters. $Ho^{3+}$ ions are doped in the $NaGdF_4$ core with a doping ratio of $x\%$ ($x$=5, 20, 100) to form the Ho$x$ cores. An inert passivation shell of $NaGdF_4$ is then epitaxially grown to form $NaGd_{1-x\%}F_4$:$Ho_{x\%}$@$NaGdF_4$ (referred to as Ho$x$@Gd) core-shell nanoparticle (NP). Upconversion is measured in a home-built microscope system delivering tunable infrared light from an optical parametric oscillator laser (details in Methods), focused onto a thin film of Ho-doped LnNPs (HoNPs) on glass using a reflective objective, and collected in transmission via another objective lens to a spectrometer for detection (Fig. 1b). Figure 1c shows bright red emission from the Ho100@Gd NPs, measured on a Si CCD camera, under the excitation of 2 µm light. Figure 1d shows that compared to previous self-sensitized LnNPs, our HoNPs demonstrate the longest excitation wavelength and the largest anti-Stokes shift [5,21-23]. Absorption of SWIR light by Ho100@Gd NPs occurs in a broad band from 1860 to 2080 nm (Fig. 1e), with other clearly assigned transitions[24] in the visible and NIR absorption of

Ho$^{3+}$ ions. The dominant red upconversion at 640 nm is assigned to the $^5F_5 \rightarrow {}^5I_8$ transition (Fig. 1f). Weak emission is also observed at 540 nm ($^5S_2+{}^5F_4 \rightarrow {}^5I_8$) and 745 nm ($^5S_2+{}^5F_4 \rightarrow {}^5I_7$), consistent with typical Ho$^{3+}$ emission. Hence, we successfully achieve direct 2 μm upconversion to 640 nm in the self-sensitized HoNP system.

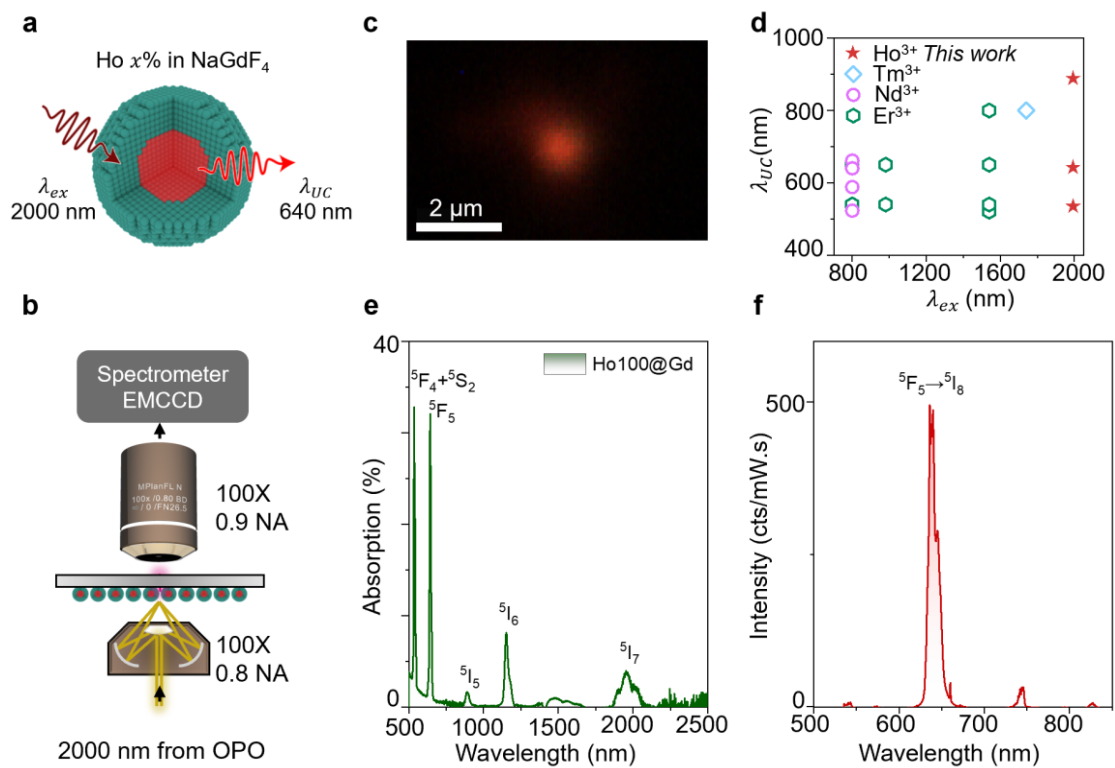

**Fig. 1 | Concept of 2 μm upconversion by Ho-based nanoconverters. a,** Schematic illustration of the core-shell nanoparticle structure of Ho$^{3+}$ doped within a NaGdF$_4$ matrix (Ho$x$@Gd) with different dopant concentrations, excited by SWIR light at $\lambda_{ex}$=2000 nm and emitting upconverted $\lambda_{UC}$=640 nm light. **b,** Schematic of upconversion dual SWIR-Visible microscope with objective lenses focusing and collecting light from HoNP films. **c,** Camera image of upconverted red light. **d,** Excitation and upconversion emission wavelength of existing self-sensitized LnNPs. **e,** Absorption spectrum of Ho100@Gd NPs with at a concentration of 100 mg ml$^{-1}$. **f,** Upconversion spectrum of Ho100@Gd NPs in a film sample, with labels indicating electronic energy levels involved in the main upconversion transition.

To investigate the effect of dopant concentration and shell passivation in self-sensitized HoNPs, we first characterized the structure of the core and core-shell HoNPs. The average sizes of Ho$x$@Gd core-shell NPs are measured to be 16.0±0.3, 15.7±0.1, 16.7±0.1 nm, for $x$=5, 20, and 100% respectively (Supplementary Fig. 1). The high-angle annular dark-field (HAADF) scanning transmission electron microscopy (STEM) image of a single Ho100@Gd NP shows a well-ordered atomic arrangement in the core (Fig. 2a). Figure 2b shows the elemental mapping using energy dispersive X-Ray spectroscopy (EDS) of a single Ho100@Gd NP, indicating a core-shell structure with Ho in the core and Gd in the shell. After denoising, a line profile across the center of the NP (Fig. 2c) shows the relative elemental distribution of Gd and Ho with a shell thickness of ~3 nm, which is consistent with the average shell thickness of ~2.5 nm measured by the TEM images of core and core-shell HoNPs (Supplementary Fig. 1). The diffraction pattern of the single core-shell NP shows a characteristic hexagonal phase, consistent with the powder X-ray diffraction pattern of Ho100@Gd NPs (Supplementary Fig. 2).

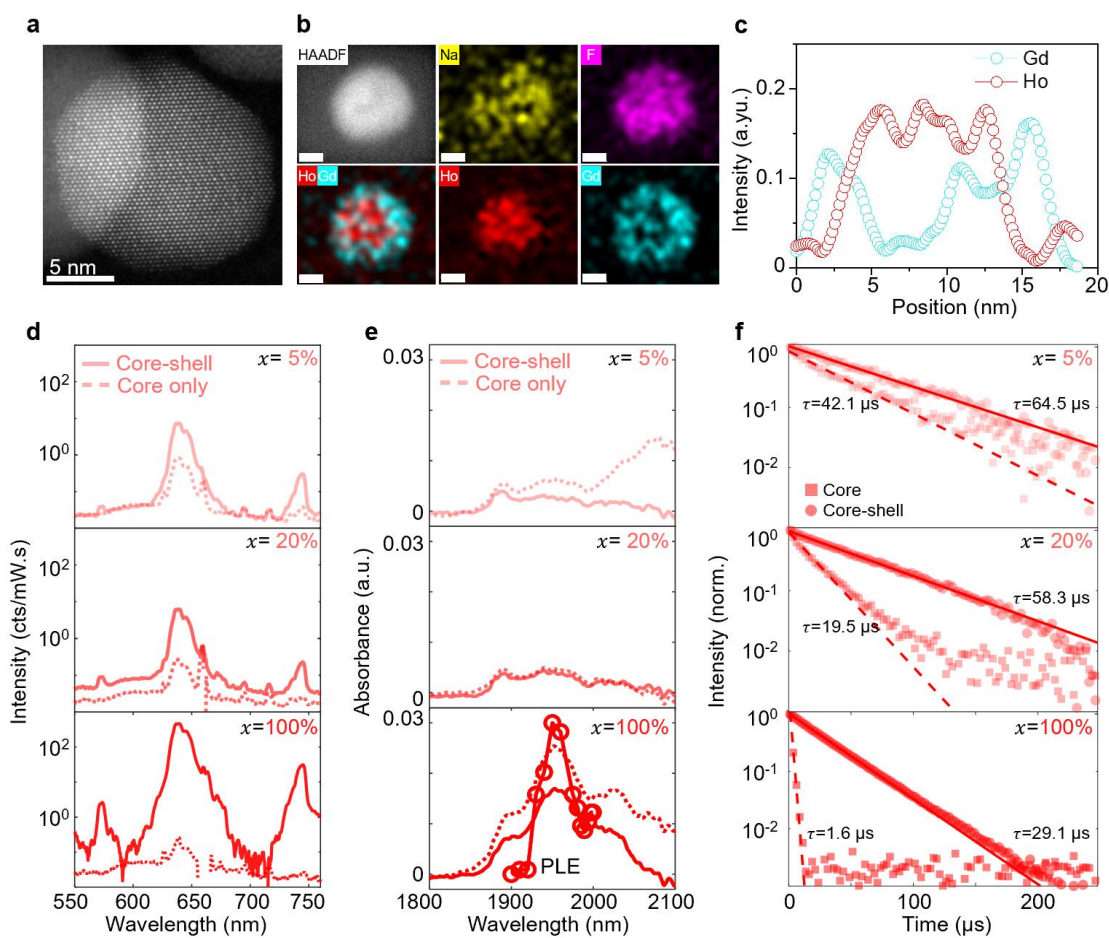

**Fig. 2 | Core-shell nanoparticles reduce surface quenching. a,** High resolution HAADF-STEM image of Ho100@Gd NP. **b,** STEM HAADF image and corresponding EDS elemental mapping of a single core-shell Ho100@Gd NP. Scale bar, 5 nm. **c,** Elemental distribution of Gd and Ho of a single NP using a line profile across the center of the NP. **d,** Upconversion PL spectra, **e,** SWIR absorption spectra, and **f,** TCSPC lifetimes of Ho$x$ cores and the corresponding Ho$x$@Gd core-shell NPs ($x$=5%, 20% and 100%). Circles at the bottom of **e** also show the upconversion photoluminescence excitation (PLE) spectrum.

We measure and compare the SWIR upconversion intensity, SWIR absorption, and emission lifetime of 640 nm before and after shelling. To ensure similar particle densities, the samples are spin-coated to form monolayers. For the Ho$x$ NPs, the upconversion intensity decreases as the dopant concentration increases (Supplementary Fig. 3). Such phenomena have been observed previously, and attributed to two reasons: concentration quenching induced by enhanced cross-relaxation between Ln ions and/or enhanced energy migration to surface defects[21,25]. After passivation with an inert shell on the cores, the upconversion intensities of core-shell NPs under 2 μm excitation has significantly improved by 10, 23, 2600-fold for $x$=5,

20, and 100%, respectively (Fig. 2d). After shelling, increasing dopant concentration results in a significant enhancement of the upconversion intensity, especially for the Ho100@Gd NPs (Supplementary Fig. 3).

Figure 2e shows the SWIR absorption of HoNPs before and after shelling. The absorbance of core HoNPs increases as the dopant concentration increases. The absorbance of core-shell NPs has slightly decreased at the same mass concentration after shelling. The photoluminescence excitation (PLE) spectrum of Ho100@Gd NPs closely matches the SWIR absorption of HoNPs. The introduction of a shell increases the emission lifetime, measured via a home-built single-NP time-correlated single-photon counting (TCSPC) microscope, for all dopant concentrations. The lifetimes of Ho$x$@Gd NPs have been prolonged to 65.4, 58.3, and 29.1 μs from 42.1, 19.5 and 1.6 μs for $x$=5, 20, and 100%, respectively (Fig. 2f). This shows that surface passivation is more effective for highly doped core NPs and energy migration to surface defects is the main photoluminescence (PL) quenching channel for highly doped HoNPs. Ho5@Gd NPs still show a longer decay lifetime than Ho100@Gd, indicating that concentration quenching by enhanced cross-relaxation can still influence the upconversion intensity but is not dominant in the highly doped systems. The combined TCSPC and HAADF-STEM data indicate that the core's surface defects are minimized, showing effective passivation and reduction of fluorescence quenching sites.

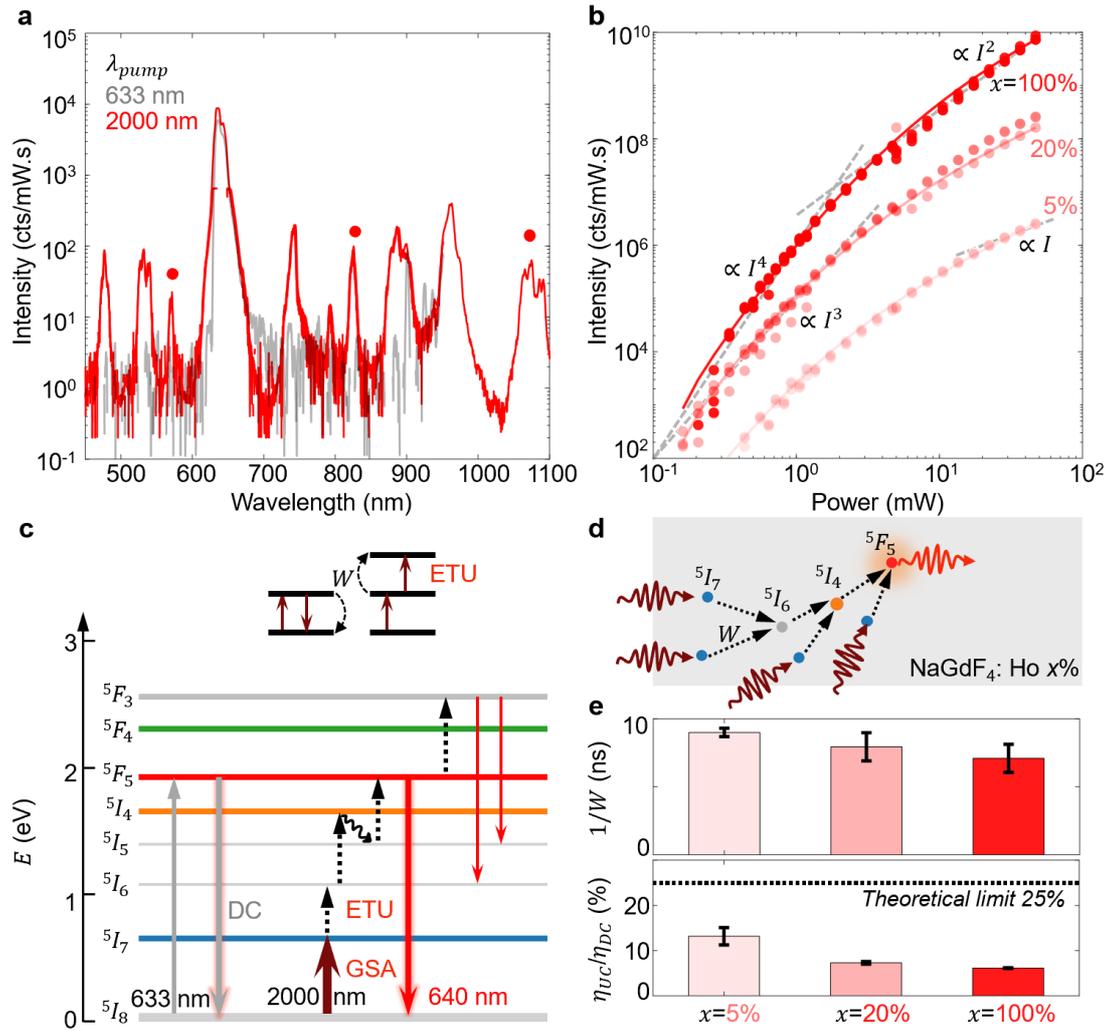

**Fig. 3 | Mechanism of efficient 2 μm upconversion. a,** Comparison of emission spectra of Ho100@Gd NPs pumped by 633 nm and 2000 nm. **b,** Pump power dependence of the SWIR upconversion intensities of Ho$x$@Gd NPs. **c,** Energy level scheme of the downconversion and energy transfer upconversion (ETU), with ETU rate $W$. Upconversion luminescent transitions are labelled with red arrows, with the main one at 640 nm and newly observed transitions labelled by red arrows. **d,** Schematic illustration of the ETU upconversion pumped by 2 μm light to generate the final emissive $^5F_5$ state from the $^5I_7$ reservoir states. **e,** Inverse of the ETU rate ($1/W$) and relative upconversion $\eta_{UC}$ vs. downconversion $\eta_{DC}$ efficiency for Ho$x$@Gd NPs ($x$=5%, 20% and 100%).

To build a deeper understanding of the upconversion process in the novel 2 μm self-sensitization system, we compare the downconversion and upconversion PL spectra pumped by 633 nm and 2 μm excitation, respectively. The PL spectra are plotted on a logarithmic intensity scale, and we find that there are 3 new peaks present in the upconverted emission (labelled in red dot in Fig. 3a). This suggests a distinct energy transfer mechanism compared to the downconversion process. The intensity of the upconversion signal is surprising, considering the energy conservation requirement of a 4-photon excitation to produce a 640 nm photon from 2000 nm photons. The instantaneous nonlinear optical process, involving a fourth-order susceptibility $\chi^{(4)}$ and virtual levels, is *a-priori* highly improbable. Here the dense electronic energy levels possessed by Ho$^{3+}$ ions are an advantage, as it provides many intermediate energy levels to shelve short-lived photon excitations as long-lived electronic excitations.

The power dependence of upconversion over four decades in intensity and three decades in incident SWIR pump power shows a highly nonlinear trend with different dopant concentrations for the core-shell Ho$x$@Gd NPs (Fig. 3b). For the Ho100@Gd NPs, the PL is initially a power law $\propto I^4$ to pump intensity $I$, which then saturates to an exponent of around $\propto I^2$. Reduced cross-relaxation pushes the saturation power to lower thresholds for the Ho5@Gd and Ho20@Gd NPs, thus the power dependence is less steep. For the Ho5@Gd NPs, the power law change is extremely striking, with the exponent decreasing from $\propto I^3$ at low powers to saturating at $\propto I$ at higher powers.

To understand the power laws and yield insight into the upconversion mechanism, we model the upconversion process with a simplified four-level scheme[26] for Ho$^{3+}$ in Figure 3c. Unlike the direct activation of the $^5F_5$ level in downconversion, the upconversion first shows a ground state absorption (GSA) and then energy transfer upconversion (ETU) process. This can be confirmed by the new emission peaks from the $^5F_3$ state. Ladder climbing from one electronic level to the next can occur via sequential photon absorption, known as excited state absorption (ESA), or via ETU. ETU occurs via Forster/Dexter energy transfer between two adjacent donor/acceptor Ln ions, the acceptor is promoted to a higher energy level with energy difference equal to that lost in the donor (Fig. 3c). Assuming a minimal steady-state model, where states decay radiatively at the measured lifetime (from Fig. 2f, indirectly accounting for non-radiative relaxation and cross-relaxation), and all energy levels perform ETU with equal rates, we show that rather than ESA, ETU is primarily responsible for the efficient upconversion we see (details in Supplementary Information Section S1).

Upconversion proceeds via the excitation of the $^5I_8$ ground state to a $^5I_7$ reservoir with the SWIR pump. This long-lived reservoir can then exchange energy via self-sensitization to excite the $^5I_6$ state, which can then capture another $^5I_7$ donor state to be excited to $^5I_4$, which rapidly relaxes to $^5I_5$, and then excited again to finally reach $^5F_5$, the terminal emitter state. Mathematically modelling this process[26] yields a good fit to the power dependence measured (solid lines in Fig. 3b). The model estimates the ETU rate $W$, the inverse of which is the average self-sensitization time (Fig. 3e). This is most efficient for the Ho100@Gd, with $1/W$ = 7.0±1.0 ns (compared to Ho20@Gd at 8.0±1.0 ns and Ho5@Gd at 8.9±0.5 ns), due to the proximity of the ions and both Forster/Dexter energy transfer rates increasing with reduced distance between donor and acceptor.

The SWIR upconversion process is highly efficient, evaluated by the ratio between the downconversion ($\eta_{DC}$) and upconversion ($\eta_{UC}$) PL quantum efficiency (Fig. 3e, details in Supplementary Information Section S2). In the limit of perfect upconversion efficiency, the upper-bound ratio of $\eta_{UC}/\eta_{DC}$ is 25%, as energy conservation requires 4 SWIR photons for every 640 nm photon produced. The $\eta_{UC}/\eta_{DC}$ ratio drops from 13.2±2.0% to 7.5±0.5% to

6±0.5%, when the Ho concentration changes from 5% to 20% to 100%. For the Ho5@Gd NPs, with the least cross-relaxation and non-radiative quenching, and longest emission lifetime (Fig. 2f), we obtain the highest $\eta_{UC}/\eta_{DC}$ efficiency = 13.2±2.0 %, which is half of the theoretical maximum. This manifests as a strong saturation in the upconversion PL power dependence, where the pump power dependence appears to be linear $\propto I$ even for this highly nonlinear upconversion process. In the limit of high efficiency harmonic generation, the power dependence is indeed predicted to become linear[27,28], although this is rarely observed due to typical conversion efficiencies being much lower than 1%[29]. The brightness of HoNPs is determined by the product of absorption and upconversion quantum yield (UCQY). Consider the significantly enhanced absorption and slightly reduced UCQY, Ho100@Gd NPs thus show the highest upconversion brightness under 2 μm excitation.

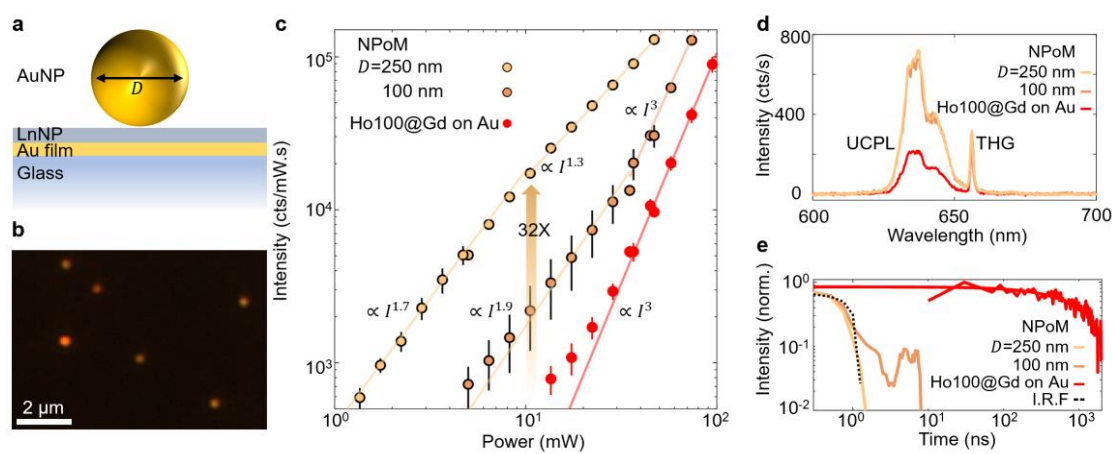

**Fig. 4│ Nanocavity enhancement of SWIR upconversion. a,** Schematic illustration of the nanoparticle-on-foil (NPoF) nanocavity constructed from gold nanoparticles (AuNP) of diameter $D$=100 and 250 nm, on top of lanthanide nanoparticle (LnNP) monolayer on a 15 nm thick Au film on a silica dielectric substrate. **b,** Dark-field microscopy image of the NPoF nanocavities. **c,** Pump power dependence of the SWIR upconversion intensity, labels indicate the power law fit to the intensity. **d,** Upconversion spectra of NPoF using 250 nm, 100 nm, and and reference Ho100@Gd monolayers on Au films. **e,** Time-resolved single photon single-particle counting microscopy for NPoF constructs, reference Ho100@Gd films, and the instrument-response function (I.R.F.).

The weak absorption, low photoluminescence quantum yield (PLQY), and long emission lifetimes of LnNPs are key limitations for applications[30]. These hurdles can be overcome by placing LnNPs within cavities resonant with their optical transitions. By selectively enhancing the out-coupling of the main upconversion emission wavelength at 640 nm, we improve the upconversion intensity. Previous NIR upconversion enhancement used dielectric metasurfaces[31,32], plasmonic nanoparticle gratings[33] and nanocube-on-mirror[34] geometries but lacked single nanocavity studies. Instead, we use the recently developed nanoparticle-on-foil (NPoF) geometry[35], which has a high field enhancement via a metal-insulator-metal-insulator (MIMI) mode[36,37]. The MIMI mode tightly confines light in the nano-sized gap between the gold nanoparticle (AuNP) and the thin gold foil. This enables emission into a dielectric waveguide[36,37] substrate. We fabricate the NPoF cavity as a proof-of-concept demonstration to enhance the upconversion performance[20,38].

The NPoF nanocavity is assembled by spin coating a monolayer of LnNPs onto a thin 15 nm Au film on a dielectric substrate, and subsequently depositing AuNPs of different diameters ($D$) to form a nanocavity between the Au metal interfaces (Fig. 4a,b). The resonance can be tuned by changing the $D$ of the AuNPs. We observe a significant upconversion enhancement within resonant NPoF cavities (Fig. 4b). The power dependence of NPoF upconversion for Ho100@Gd NPs shows a saturated response $\propto I^2$ at low powers, relative to the monolayer LnNP reference on the thin Au film (Fig. 4c). At an intermediate power, the relative enhancement can be as high as 32-fold between the bare LnNPs and the NPoF nanocavity with AuNPs of diameter $D$=250 nm. The relative enhancement is higher for the $D$=250 nm compared to 100 nm nanocavities, as the $D$=250 nm nanocavity is more on-resonant with the emission transition and excitation. The upconversion spectrum shows a similar spectrum to the reference film, but with all nonlinear optical processes also enhanced, including the instantaneous third-harmonic generation (Fig. 4d). The mechanism of intensity enhancement is mainly a Purcell enhancement of the emission, shown by the emission lifetime ($\tau$) reduction down to the instrument-response within the nanocavity ($\tau_{NC}$ < 1 ns) vs. outside ($\tau_0$ = 29.1 μs) (Fig. 4e). This yields a Purcell enhancement factor ($F_P = \tau_0/\tau_{NC}$) of > 3x10$^4$. Such high Purcell enhancement factors, better than the state-of-the-art[39] is only possible due to the tightly confined nanogap modes in a NPoF geometry.

The nanocavity increases the upconversion intensity by resonant plasmonic enhancement of the 640 nm emission. This method is universal and can be applied to improve the low emission brightness of numerous existing LnNP systems. In addition to emission enhancement, future work is needed to explore the possibility of enhancing the absorption of HoNPs by matching the NPoF resonance to the 2000 nm band. This will open further opportunities in the design and improvement of nanocavities in the SWIR range. Future optimization of the nanocavity should take advantage of designing resonances in the visible and SWIR region simultaneously[40,41], targeting both absorption and emission enhancement. Integrating the current nanocavity enhanced SWIR upconversion system into a waveguide device will also open up new applications in integrated silicon photonics.

In conclusion, we have demonstrated an efficient 2 μm upconversion system by self-sensitized Ho-doped nanoconverters to generate 640 nm emission. This system has expanded the excitation wavelength of LnNPs to the SWIR range and achieved the longest excitation wavelength and largest anti-Stokes shift to date. We showed this upconversion is a 4-photon conversion process with an ETU upconversion mechanism. The dopant concentration tuning and core-shell structure fabrication have significantly enhanced the SWIR upconversion brightness while maintaining high relative UCQYs. The overall relative efficiency of the SWIR upconversion is remarkably high, up to half of the theoretical maximum, due to highly efficient ETU between self-sensitized Ho$^{3+}$ ions. Placing a monolayer of Ho100@Gd NPs inside NPoF nanocavities further enhances the upconversion intensity by 32-fold and significantly shortened emission lifetimes to below 1 ns. Our Ho-based nanoconverters provide a new series of materials to efficiently convert SWIR light to visible light and unlock new applications of SWIR light in quantum photonics, bio-imaging, and low-loss optical communications.

**Methods**

**Lanthanide-doped nanoparticle synthesis and purification.**

The synthesis of core and core-shell HoNPs was carried out using a modified thermal decomposition method. The core Ho$x$ NPs were synthesized using overall 1 mmol of Ln(CH$_3$COO)$_3$ precursor, with $x$% Ho(CH$_3$COO)$_3$ and 1-$x$% Gd(CH$_3$COO)$_3$ dissolving in 6 mL oleic acid and 14 ml 1-octadecene with magnetic stirring in a 50 mL round bottom flask. The precursor was kept at 140 °C for 1 hr under the protection of N$_2$ flow. The precursor was cooled down to room temperature and then added a 7 mL methanol solution containing 0.07 g NaOH and 0.1 g NH$_4$F. The mixture was kept at 70 °C for 50 min to get rid of methanol. After this, the flask was moved to a heating mantle and increased the reaction temperature to 300 °C and kept at this temperature for 45 min. The solution was washed by adding ethanol and centrifuged at 7500 rpm for 5 min. The core HoNPs were precipitated at the bottom of the centrifuge tube. The core HoNPs were then dissolved in 4 mL of hexane. The shell precursor is prepared using the same method but using 0.8 mmol of Gd(CH$_3$COO)$_3$. 2 mL of the core HoNPs was added along with a 4 mL methanol solution containing 0.04 g NaOH and 0.06 g NH$_4$F. The mixture was kept at 70 °C for 45 min and then transferred to heating mantle to react at 300 °C for 45 min. The Ho$x$@Gd NPs were centrifuged and washed with ethanol for several times. All the NPs should be purified by low-speed centrifuge at 4000 rpm to get rid of large NPs or aggregated NPs.

**Nanoparticle structural characterization.**

The UV-Vis-NIR-SWIR absorption of all the NPs were carried out on Cary 7000 Universal Measurement Spectrophotometer (UMS) measuring the transmission of solution samples. The TEM images were taken on FEI Tecnai Osiris 80-200. The acceleration voltage for the electron beam was set to 200 kV. STEM HAADF and STEM EDS were collected on Titan Spectra 300 (S) TEM. The acceleration voltage for the electron beam was set to 300 kV. STEM-EDS spectrum images were acquired with a beam current of 20 pA with 300 Frames and a camera length of 115 mm. The spectrum was then denoised using principal component analysis with HyperSpy 1.7 for line profile analysis.

**Nanocavity (NPoF) fabrication and upconversion monolayer preparation.**

Monolayers of core-shell Ho@$x$Gd and core-only Ho$x$ NPs were prepared by adding Ho@$x$Gd solution (25 mg mL$^{-1}$ in hexane) to ethanol to precipitate the NPs. The resulting suspension was centrifuged at 9500 rpm for 5 minutes to separate the precipitate from the supernatant, which was then discarded. The sample was then weighed and resuspended in octane to 25 mg mL$^{-1}$ concentration with vortex mixing and sonication. HoNPs in octane were then spin coated to form monolayers on various substrates at 2000 rpm, 2000 rpm/s acceleration, for 30s onto various substrates. The spin-coated layer was then annealed at 90°C for 1 minute. For the upconversion efficiency measurements, they were spin coated onto a glass coverslip cleaned by rinsing for 1 minute with water, acetone, and isopropanol before drying in nitrogen gas, and then oxygen plasma cleaned using 90% RF power & 30 cm$^3_{STP}$ min$^{-1}$ (Henniker Plasma, HPT-100) for 1 min. Nanoparticle-on-foil (NPoF) fabrication was performed by spin coating a monolayer onto a thin layer of gold (15 nm) which was thermally evaporated (Kurt Lesker, 0.5 Å/s rate) onto a thin glass coverslip (cleaned by rinsing for 1 minute with water, acetone, and isopropanol before drying in nitrogen gas). The layer of gold was then oxygen plasma cleaned using 90% RF power & 30 cm$^3_{STP}$ min$^{-1}$ (Henniker Plasma, HPT-100) for 1 min prior to spin-coating. After deposition, the monolayer was annealed at 90°C for 1 minute prior to nanoparticle deposition. Citrate-capped gold nanoparticles (100 nm, and 250 nm diameter, BBI Solutions, first suspended in a 1:5 volume ratio with 1M aqueous NaCl solution) were deposited by drop casting solutions onto the lanthanide monolayer on Au film and waiting 30 s before washing off the droplet with de-ionised water and cleaning with nitrogen gas. The assembled NPoF structures were then verified with dark-field microscopy and spectroscopy.

**SWIR excitation, photoluminescence, and dark field spectro-microscopy.**
Optical spectra were recorded with custom-built dual mid-infrared and visible dark field microscope with a motorized stage and spectrometers. The setup is fully automated to record the optical spectra of many plasmonic nanocavities. Particle recognition algorithms are used to locate NPoF cavities by their dark field scattering and centre them to the measurement spot. A NIR pump (1032 nm, 2 ps, 80 MHz, 16 W, Emerald Engine IR HP) is sent to an optical parametric oscillator (APE, Levante IR) to generate a SWIR signal beam (1300-2000 nm). The SWIR beam is then sent to a home-built microscope via a Cassegrain objective lens (Pike, 100X, 0.8 NA) and focused down to a spot with a ~2 µm full-width-at-half-maximum. The generated upconversion luminescence is then collected with a dark-field objective lens (Olympus, 100X, 0.9 NA) in transmission. The SWIR nm pump is then filtered-out and the luminescence spectrum detected via a spectrometer (Kymera, 328I) through a 200 µm wide slit which disperses the signal with a 150 lines/mm grating onto a CCD camera (Andor Newton920). Downconversion measurements were performed with a 633 nm pump laser (Matchbox) coupled into the same microscope. Dark field (DF) scattering spectra were recorded with incoherent white light (halogen lamp) through an Olympus MPLFLN 100x DF objective with 0.9 NA and recording the scattered light with a fibre-coupled spectrometer (Ocean Optics QE Pro).

**Single-particle time-resolved single-photon counting microscopy.**
Time-resolved photoluminescence was recorded on the home-built microscope with a 640 nm pulsed excitation source (PicoQuant LDH-P-C-400B, driven at 4kHz, 40 kHz and 40MHz to obtain different timescales, filtered by an angle-tuned 650 nm bandpass to reduce the laser spectral side-wings). The luminescence was collected with a darkfield objective lens (Olympus, 100X, 0.9 NA) in reflection and filtered with a 650 nm long-pass before routing to a single-photon avalanche photodiode (Micro Photon Devices PDM PD-100-CTD). The arrival times of all photons were continuously recorded with a time-tagged system on a field-programmable gate array board (details in [42]).

**Data availability**
The data in the main text and Supplementary Information are publicly available from the University of Cambridge repository.

**Acknowledgements**
Z.Y. acknowledges funding from UKRI Postdoctoral Individual Fellowships (Grant Reference EP/X023133X/1). R.A. acknowledges support from the Winton Programme for the Physics of Sustainability and from St. John's College Cambridge. The authors acknowledge funding from the EPSRC (EP/L027151/1 and EP/R013012/1), and the ERC (883703 PICOFORCE). The authors thank Prof. Jeremy Baumberg for experimental facilities in the Cambridge NanoPhotonics Centre, and Prof. James P. Schuck (Columbia University) for discussions.

**Author contributions**

Z.Y. conceived the idea. Z.Y. and R.A. designed the experiments. Z.Y. performed nanoparticle synthesis and purification and carried out UV-Vis-NIR-SWIR absorption and XRD measurement. R.A. performed all upconversion measurement, designed and fabricated the cavity system, analyzed upconversion spectra and performed theoretical modelling. F.M.B. contributed to the upconversion measurements. Z.J. performed TEM measurement. X.L. performed STEM measurement for elemental mapping under the supervision of C.D. A.T. contributed to the upconversion mechanism. Z.Y., R.A. and A.R. wrote the manuscript. Z.Y. and A.R. supervised the project. All authors discussed the results and commented on the manuscript.

**Competing interests**

The authors declare no competing interests.